\documentclass[conference]{IEEEtran}

\ifCLASSINFOpdf

\else

\fi

\usepackage{cite}
\usepackage{amsmath,amssymb,amsfonts}
\usepackage{algorithmic}
\usepackage{graphicx}
\usepackage{textcomp}
\usepackage{xcolor}
\begin{document}

\title{Efficient Deep Learning Adaptation for Cross-Environment RSS-Based Indoor Localization
}

\author{Cien~Zhang\IEEEauthorrefmark{1}, Jiaming~Zhang\IEEEauthorrefmark{2}\\
\IEEEauthorrefmark{1}Wharton Research Data Services, University of Pennsylvania, Philadelphia, USA\\
\IEEEauthorrefmark{2}Department of Electronic, Electrical and Systems
Engineering, University of Birmingham, Birmingham, U.K.
}

\markboth{Journal of \LaTeX\ Class Files,~Vol.~14, No.~8, August~2015}%
{Shell \MakeLowercase{\textit{et al.}}: Bare Demo of IEEEtran.cls for IEEE Journals}

\maketitle

\begin{abstract}Received signal strength (RSS)-based indoor localization has attracted increasing attention due to its low cost and compatibility with existing wireless infrastructures. However, RSS measurements are highly sensitive to environmental variations, making it challenging for deep learning-based localization models to generalize across different physical configurations. Driven by these challenges, this paper proposes a scenario-adaptive RSS localization framework based on backbone reuse and efficient adaptation. The proposed architecture consists of a lightweight extractor and a shared backbone, where the extractor projects heterogeneous RSS inputs into a unified feature space and the backbone captures transferable localization knowledge. During adaptation, the backbone trained from the source dataset is reused, while the extractor is optimized to adapt to the new dataset configuration. A short optimization stage is further introduced to slightly refine the backbone with a lower learning rate. Experimental results on four datasets demonstrate that the proposed training strategy enables faster convergence and improved adaptation compared with training from scratch. In addition, the proposed framework reduces training time under different dataset configurations, verifying its effectiveness and efficiency for adaptive RSS-based indoor localization.
\end{abstract}

\begin{IEEEkeywords}
Deep learning, indoor localization, received signal strength
\end{IEEEkeywords}

\section{Introduction}
\label{sec:introduction}
\subsection{Background}
\IEEEPARstart{I}{ndoor} positioning systems leveraging received signal strength (RSS) have become
ubiquitous, primarily because they can be seamlessly deployed over existing Wi-Fi or
Bluetooth networks without requiring specialized hardware. These advantages make it suitable for a variety of applications, including navigation, target tracking, and surveillance \cite{park2022multidirectional, yin2017distributed,he2026rss}. However, compared with time-based and angle-based localization techniques, RSS-based methods generally exhibit lower accuracy. This limitation mainly arises from the sensitivity of RSS measurements to environmental variations, such as multipath propagation and channel fading \cite{11458632}.

To mitigate these issues, fingerprinting-based localization has been widely investigated\cite{9542942, 11397142}. By establishing a mapping between RSS observations and corresponding spatial locations, fingerprinting approaches can achieve reliable positioning performance even in complex indoor environments with severe non-line-of-sight (NLoS) conditions \cite{10476652}. Fingerprinting-based localization typically involves two key stages: (1) an offline stage and (2) an online stage. In the offline stage, a fingerprint database is constructed by collecting RSS measurements at predefined reference points. In the online stage, newly observed RSS data are compared with the stored fingerprints to infer the target position. Although this framework can provide satisfactory localization accuracy, several practical challenges still need to be addressed.

Indoor localization has been extensively studied in recent years, with existing approaches broadly categorized into range-based and fingerprint-based methods. Range-based techniques estimate distances between reference points and targets using measurements such as time of arrival (ToA) or angle of arrival (AoA), followed by triangulation or trilateration for position estimation. However, these methods often require strict line-of-sight conditions and precise synchronization, which limit their applicability in complex indoor environments. In contrast, fingerprint-based approaches leverage received signal strength (RSS) measurements to learn a mapping between signal features and spatial locations, demonstrating superior robustness under non-line-of-sight (NLoS) conditions. Nevertheless, such methods typically rely on labor-intensive site surveys to construct fingerprint databases, which hinders scalability in large-scale or dynamic environments.

\subsection{Related Work}

To improve localization performance, recent studies have explored the use of deep learning techniques for RSS-based positioning \cite{lin2026variational, 11362365, 7366598}. By learning complex nonlinear relationships between RSS measurements and spatial coordinates, deep learning models have substantially reduced localization errors. In particular, deep learning models can better capture the underlying structure of wireless signals. However, most existing deep learning-based approaches are trained and evaluated within a fixed environment, and their performance degrades significantly when deployed in new scenarios due to the distribution shift in RSS measurements. 

To address this issue, transfer learning (TL) has been introduced into indoor localization to enable knowledge transfer across different domains. For instance, \cite{8030076} addresses domain shift in RSS-based localization by transferring distance metrics from source to target domains to reshape the feature distribution for clustering-based localization. While effective, these methods are typically built upon handcrafted features and clustering-based frameworks, limiting their representation capacity and adaptability. \cite{10274764} proposed a meta-learning-based framework, MetaLoc, for indoor localization, which enabled fast adaptation across different environments within a similar physical setup through transferable initialization. However, it relied on iterative fine-tuning of the entire model and is mainly evaluated on tasks generated under consistent data collection settings, where domain discrepancies are relatively limited. \cite{11026791} proposed a meta-learning-based two-part framework for channel state information (CSI)-based localization, separating environment-independent and environment-specific components to improve adaptation. It evaluates cross-environment performance using multiple scenarios from a unified dataset, Dichasus \cite{dichasus2021}, where different base stations are treated as separate environments. However, it assumes aligned input structures (e.g., via subarray selection) and relies on fine-tuning strategies such as gradual unfreezing, with domain discrepancies remaining relatively limited compared to cross-dataset settings.

\subsection{Contribution}

To address the limitations of existing methods, this work proposes a scenario-adaptive localization framework that enables efficient knowledge transfer across environments. The main contributions are summarized as follows:

\begin{enumerate}
    \item A modular architecture is proposed, consisting of an extractor and a shared backbone, to decouple feature alignment from localization learning. The backbone is designed to capture scenario-invariant knowledge, while lightweight adaptation is performed through a small set of parameters to accommodate scenario-specific variations.
    
    \item An efficient adaptation strategy is developed, where the backbone is primarily kept frozen to preserve transferable knowledge. In addition, a short and controlled refinement stage is introduced, during which the backbone is selectively unfrozen for a limited number of epochs. This step aims to enhance representation generalization, thereby improving robustness across different scenarios while maintaining low adaptation cost.
    
    \item Extensive experiments demonstrate that the proposed method achieves robust localization performance under various domain shifts, while maintaining high efficiency in adaptation.
\end{enumerate}

\noindent \textit{Notation}:
Bold lower-case letters represent column vectors, while capital letters are matrices. The superscripts $(\cdot)^T$ and $(\cdot)^{-1}$ represent the transpose and inverse operations, respectively. The estimate of the variable $x$ is denoted by $\hat{x}$, $\|\cdot\|$ is the Euclidean norm, while $\odot$ denotes element-wise multiplication.

\section{Problem Formulation}
\label{sec:Problem Formulation}

\subsection{RSS Measurement}

We consider a two-dimensional (2-D) localization problem, where the objective is to estimate the target position $\mathbf{p} = [x \ y]^T$ using $L$ radio units (RUs). Let $\mathbf{p}_l$ denote the location of the $l$-th RU, and $P_t$ represent the transmit power of the target device. The received signal power at the $l$-th RU can be modeled as \cite{he2026rss, so2011linear}:
\begin{equation}
    P_{r,l} = P_t K_l h_l d_l^{-\alpha} w,
\end{equation}
where $d_l = \| \mathbf{p} - \mathbf{p}_l \|$ denotes the Euclidean distance between the target and the $l$-th RU, $K_l$ captures antenna gain and other non-propagation-related effects, $h_l$ represents the channel fading gain, and $\alpha$ is the path-loss exponent. The term $w$ models large-scale shadowing effects and is typically assumed to follow a log-normal distribution.

By taking the natural logarithm, the received signal strength (RSS) can be expressed as
\begin{equation}
    r_{\text{RSS},l} = -\alpha \ln d_l + n,
    \label{eq:rss}
\end{equation}
where $n$ represents the aggregated noise term, commonly approximated as zero-mean additive white Gaussian noise (AWGN). 

In practical environments, however, RSS measurements are highly unstable due to multipath propagation, shadowing, and Non-Line-of-Sight (NLoS) conditions. These factors introduce significant uncertainty and nonlinearity into the measurement model, which severely degrades the performance of conventional multilateration-based localization methods \cite{9122501,11165772, 10077563}. 

\subsection{Fingerprinting Localization}

Consequently, fingerprinting has emerged as a robust data-driven alternative. Instead of relying on an explicit propagation model, fingerprinting directly learns the mapping between RSS measurements and spatial locations.

The fingerprinting framework consists of two phases:

\textit{1) Offline Phase:} The localization area is partitioned into $I \times J$ grids with spatial resolution $\epsilon$. The centre of each grid is defined as a reference point (RP), denoted by $\mathbf{p}_{ij} = [x_{ij}, y_{ij}]^T$. At each RP, RSS samples are collected from all RUs to construct the fingerprint database. Specifically, the RSS vector observed at the $l$-th RU is given by
\begin{equation}
    \mathbf{r}_{\text{RSS},i,j,l} = \left[r^{1}_{\text{RSS},i,j,l}, r^{2}_{\text{RSS},i,j,l}, \cdots, r^{N}_{\text{RSS},i,j,l}\right]^T,
\end{equation}
where $N$ denotes the number of RSS measurements, which may include both Line-of-Sight (LoS) and NLoS conditions.

\textit{2) Online Phase:} Given a newly observed RSS vector, pattern matching is performed to estimate the target location by identifying the RP whose stored fingerprint is most similar to the observation. Common similarity metrics include cosine similarity and other kernel-based measures.

Although fingerprinting improves robustness in complex propagation environments, it suffers from a fundamental limitation in handling environmental dynamics. In practice, changes in propagation conditions, particularly variations in physical configurations and sensing setups, such as the number and placement of receivers, can significantly alter the RSS distribution. As a result, the fingerprint database constructed in the offline phase often becomes outdated when deployed in a new environment, leading to severe performance degradation.

A straightforward solution is to retrain the localization model for each new scenario. However, this process is highly labor-intensive and computationally expensive, making it impractical for large-scale or real-time applications. Moreover, existing methods typically lack the ability to efficiently transfer knowledge across different physical environments, resulting in limited generalization capability under domain shifts.

These limitations highlight the need for more advanced approaches that can effectively exploit the statistical and spatial characteristics of RSS data, while enabling efficient adaptation across varying scenarios without requiring extensive retraining or dense data collection.

\section{Proposed Method}
\subsection{Model Architecture}
\begin{figure}
    \centering
    \includegraphics[width=0.7\columnwidth]{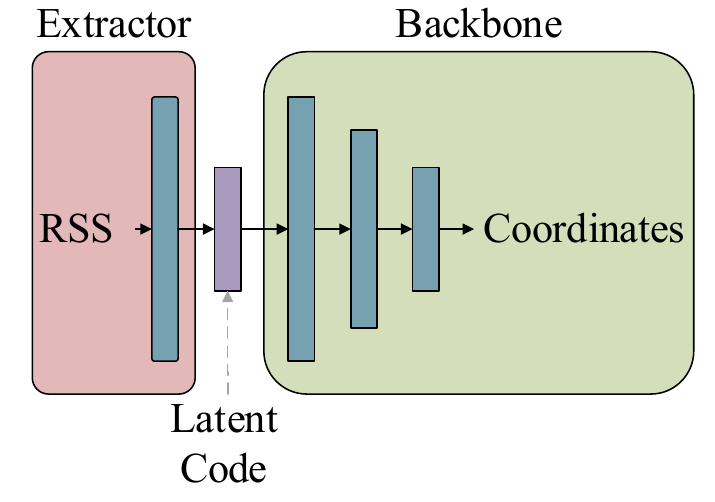}
    \caption{The architecture of the proposed deep learning model. It consists of the extractor and the backbone. The input of the extractor is the RSS data, and the output of the backbone is the coordinates.}
    \label{fig:architecture}
\end{figure}

The motivation of the proposed method is to design a method enabling a clear separation between feature alignment and localization learning, allowing the model to generalize across scenarios while requiring only minimal adaptation, which is particularly beneficial in practical deployments with limited labelled data. As shown in Fig. \ref{fig:architecture}, the network follows a modular architecture composed of an extractor and a backbone. 

In particular, the extractor, implemented as a fully connected layer, serves a dual purpose. First, considering the number of receivers used in different physical environments could be different, it normalizes heterogeneous input representations by projecting RSS measurements of varying dimensions into a unified feature space by applying a fully connected layer. This fully connected layer also performs preliminary feature embedding, supporting the backbone to capture essential signal characteristics. Built upon this unified representation, the backbone network consisting of three fully connected layers acts as the core localization module. It learns the mapping from the extracted features to spatial coordinates, while maintaining robustness across different environmental conditions. This decoupled design facilitates efficient domain adaptation: the backbone captures generalizable localization knowledge, while the lightweight extractor can be selectively adapted to accommodate scenario-specific variations, significantly reducing retraining cost and improving scalability.

As observed, the proposed model adopts fully connected layers only. The rationale behind this design is that the RSS measurements are represented as global feature vectors rather than spatially correlated image data. Unlike natural image processing tasks, the reconstruction problem mainly involves learning a global non-linear mapping between the measurement space and target space. Therefore, lightweight fully connected architectures are sufficient to capture the underlying relationships without relying on more computationally intensive operations such as convolutions.

In addition, compared with large-scale deep learning models such as Transformers or foundation models, the proposed lightweight architecture offers lower computational complexity, reduced training cost, and improved stability under limited radar datasets. Nevertheless, for more challenging scenarios involving larger-scale datasets, stronger environmental variations, or more complicated target structures, advanced architectures could be further explored to enhance model representation capability and generalization performance.

\subsection{Training Strategy}
\begin{figure}
    \centering
    \includegraphics[width=\columnwidth]{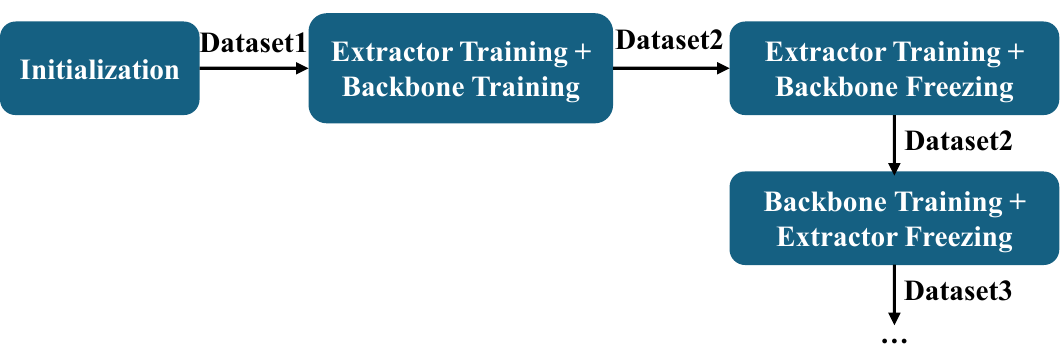}
    \caption{The training flow of the proposed model.}
    \label{fig:WorkingFlowoftheOptimization}
\end{figure}

In this work, the training strategy plays an important role, and the overall workflow is illustrated in Fig. \ref{fig:WorkingFlowoftheOptimization}. First, the proposed model is trained using the collected dataset reported in \cite{zhang2025fingan}, referred to as Dataset1. During this stage, both the extractor and the backbone are jointly trained in an end-to-end manner, enabling the proposed network to achieve an initial optimization.

Subsequently, the model is further optimized using another dataset, referred to as Dataset2. Different from the training process with Dataset1, the optimization with Dataset2 mainly focuses on the extractor, while the trainable parameters of the backbone are frozen. In this way, the influence of backbone parameter variations can be avoided during the optimization of the extractor, allowing the extractor to better adapt to the new dataset characteristics.

After the extractor training is completed, an additional optimization stage is introduced to further improve the robustness of the backbone. Specifically, the backbone is fine-tuned using a lower learning rate while keeping the extractor parameters fixed. This strategy enables the backbone to achieve slight adaptive refinement without affecting the optimized feature extraction capability of the extractor.

Therefore, the proposed staged optimization strategy enables the network to gradually improve its adaptation capability and robustness under different dataset characteristics while maintaining the stability of the learned feature representations.

\subsection{Experimental Details}
The proposed model is implemented in Python 3.6 using TensorFlow, and the network is trained and evaluated on an NVIDIA Tesla A40 GPU. Four different datasets are selected in this work, and the corresponding training/testing splits and sample numbers are summarized in Table \ref{tab:datasets}. The number of filters in the extractor is set to 128, while the numbers of filters in the backbone are 64, 16, and 2, respectively.

The localization performance is evaluated using the root mean squared error (RMSE), which is defined as follows:
\begin{equation}
\text{RMSE} = \sqrt{\frac{\sum\limits_{t=1}^{T}(\hat{x}{t} - x{t})^{2} + (\hat{y}{t} - y{t})^{2}}{T}}.
\end{equation}
where $\hat{x}{t}$ and $\hat{y}{t}$ represent the predicted coordinates along the x-axis and y-axis, respectively, while $x_{t}$ and $y_{t}$ denote the corresponding ground-truth coordinates. $T$ denotes the total number of samples.

To train the proposed model, two Adam optimizers are sequentially employed for each dataset. During the extractor training stage, an Adam optimizer with a learning rate of $1\times10^{-4}$ is adopted for 100 epochs. Subsequently, another Adam optimizer with a reduced learning rate of $1\times10^{-5}$ is employed to further fine-tune the model for an additional 10 epochs.

\begin{table}[h]
\centering
\caption{Summary of the training and testing datasets used in this work.}
\begin{tabular}{|c|c|c|c|c|}
\hline
\textbf{Dataset}  & \textbf{Dataset1} & \textbf{Dataset2 \cite{feng2022analysis}} & \textbf{Dataset3 \cite{11016042}} & \textbf{Dataset4 \cite{yuen2022wi}} \\ \hline
\textbf{Training} & 13248                 & 1000                 & 8499                 & 22616                 \\ \hline
\textbf{Testing}  & 1996                & 1000                & 1812                & 5655                \\ \hline
\end{tabular}
\label{tab:datasets}
\end{table}

\section{Results}
\begin{figure}[h]
    \centering
    \includegraphics[width=\linewidth]{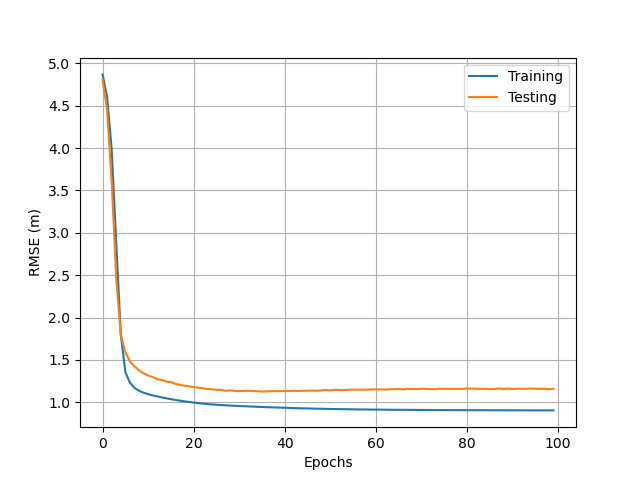}
    \caption{Training and testing curves using Dataset1.}
    \label{fig:dataset1}
\end{figure}

The training and testing results with Dataset1 is shown in Fig. \ref{fig:dataset1}. As observed, both the training and testing curves tend to be stable when it is $100^{th}$ epoch. During the initial training stage, the RMSE decreases rapidly for both the training and testing datasets, indicating that the proposed model is capable of effectively learning the underlying mapping relationship between the RSS measurements and the target coordinates.

As the number of epochs increases, both curves converge to relatively stable values, demonstrating that the proposed model achieves stable convergence during the training process. In addition, the testing curve follows a similar trend to the training curve without significant fluctuation or divergence, which indicates that the proposed model maintains good generalization capability and does not exhibit obvious over-fitting. Although a small performance gap exists between the training and testing results, such behaviour is expected due to the difference between seen and unseen samples. Overall, the convergence behaviour verifies that the proposed model can effectively accomplish the localization task under the first physical configuration. Meanwhile, the backbone network learns a certain degree of localization knowledge from Dataset1, which provides a reliable foundation for the subsequent adaptive learning and cross-dataset optimization stages.

After training the proposed model with Dataset1, two different training strategies are further investigated on Dataset2. In the first strategy, which follows the procedures shown in Fig. \ref{fig:WorkingFlowoftheOptimization}, the extractor is randomly initialized, while the backbone adopts the optimized parameters obtained from the training process using Dataset1. In this way, the localization knowledge learned from the first physical configuration can be transferred to Dataset2 for adaptive learning. 

In contrast, the second strategy initializes both the extractor and the backbone randomly, such that the entire proposed model is trained from scratch using Dataset2. This strategy is used as a baseline to evaluate the effectiveness of the shared-backbone learning framework. 

\begin{figure}[h]
    \centering
    \includegraphics[width=\linewidth]{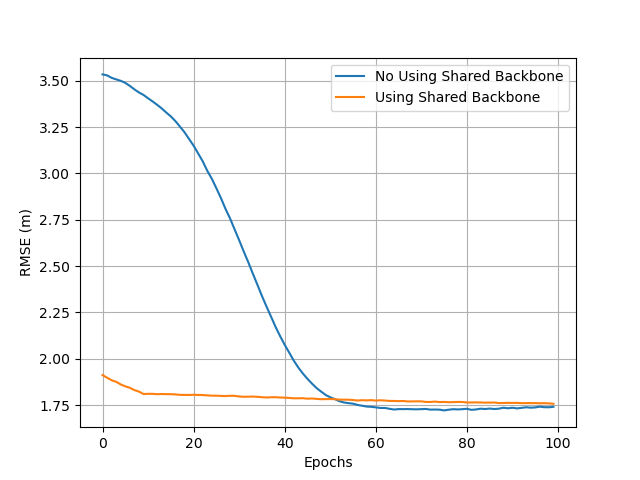}
    \caption{Comparison of the testing RMSE curves with and without the shared backbone on Dataset2.}
    \label{fig:dataset2}
\end{figure}

\begin{figure}[h]
    \centering
    \includegraphics[width=\linewidth]{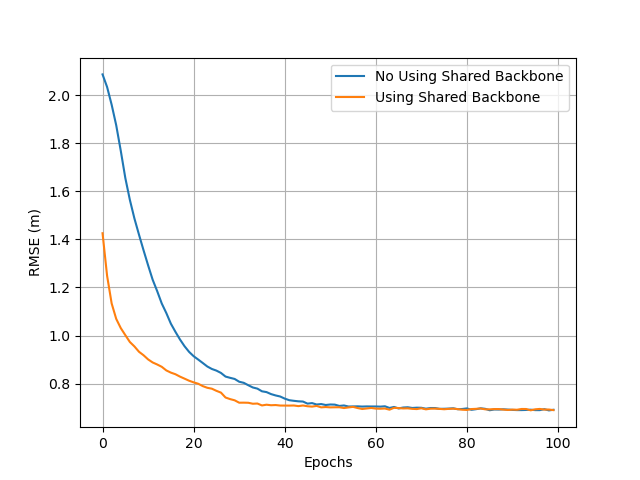}
    \caption{Comparison of the testing RMSE curves with and without the shared backbone on Dataset3.}
    \label{fig:dataset3}
\end{figure}

\begin{figure}[h]
    \centering
    \includegraphics[width=\linewidth]{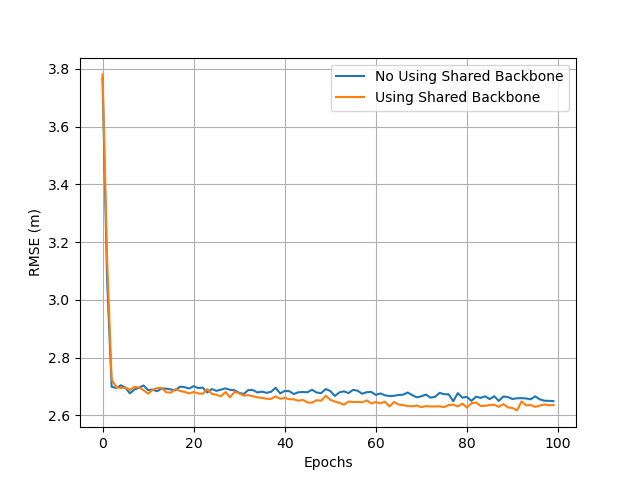}
    \caption{Comparison of the testing RMSE curves with and without the shared backbone on Dataset4.}
    \label{fig:dataset4}
\end{figure}

The results are shown in Fig. \ref{fig:dataset2}. As observed, the proposed model using the shared backbone achieves a significantly faster convergence rate compared with the model trained from scratch. In addition, the initial testing RMSE of the shared-backbone strategy is substantially lower, indicating that the backbone trained with Dataset1 has already learned a certain degree of localization knowledge from the first physical configuration. Therefore, such prior knowledge can be effectively transferred to Dataset2, enabling the proposed model to adapt to the new dataset more efficiently. In contrast, the model without the shared backbone requires more training epochs to gradually learn the localization characteristics of Dataset2. The comparison verifies the effectiveness of the proposed shared-backbone learning strategy for adaptive localization tasks across different physical configurations. 

Note that after the training of extractor, the backbone is also slightly trained for 10 epochs with a learning rate of $1\times10^{-5}$.

Similarly, the results of Dataset3 and Dataset4 are provided following the same comparison procedure. For Dataset3, a similar behaviour to that observed in Dataset2 can be identified. The proposed model using the shared backbone achieves a lower initial testing RMSE and a relatively faster convergence trend compared with the model trained from scratch, further demonstrating that the backbone is capable of transferring localization knowledge across different physical configurations.

For Dataset4, although the convergence rates of the two strategies are relatively similar, the shared-backbone strategy consistently achieves slightly lower testing RMSE values during the training process. This behaviour indicates that the shared backbone still provides beneficial prior localization knowledge, leading to lower localization errors and smoother convergence on the target dataset.

We also calculated the training time under different datasets and training strategies, and the results are summarized in Table \ref{tab:time}. As observed, the proposed training strategy requires less training time compared with the model trained from scratch, demonstrating the improved training efficiency of the proposed adaptive learning framework.

\begin{table}[h]
\caption{Comparison of training time when applying different training strategies.}
\begin{tabular}{|c|c|c|c|c|}
\hline
\textbf{Training Strategy}                                                                        & \textbf{Dataset1} & \textbf{Dataset2} & \textbf{Dataset3} & \textbf{Dataset4} \\ \hline
\textbf{\begin{tabular}[c]{@{}c@{}}Using \\ Shared Backbone (s)\end{tabular}}                     & N/A               & 20.46             & 76.94             & 186.92            \\ \hline
\textbf{\begin{tabular}[c]{@{}c@{}}Not Using \\ Shared Backbone (s)\end{tabular}} & 32.16             & 20.49             & 90.09             & 216.44            \\ \hline
\end{tabular}
\label{tab:time}
\end{table}

Overall, the experimental results obtained from Dataset2, Dataset3, and Dataset4 consistently demonstrate that the proposed shared-backbone learning framework is capable of transferring localization knowledge across different physical configurations and dataset environments. By preserving the optimized backbone obtained from Dataset1, the proposed model achieves improved initialization, faster convergence behaviour, and enhanced localization performance in most adaptive learning scenarios. These observations indicate that the backbone network is able to learn generalized localization-related feature representations rather than merely memorizing dataset-specific characteristics. Therefore, the proposed framework provides an effective solution for adaptive indoor localization tasks under different environmental configurations with reduced training complexity.

\section{Conclusion}
In this paper, a scenario-adaptive RSS localization framework was proposed to improve the adaptability and training efficiency of deep learning-based indoor localization under different physical configurations. The proposed model adopts a modular architecture consisting of an extractor and a shared backbone, where the extractor is responsible for mapping heterogeneous RSS measurements into a unified feature space, while the backbone learns transferable localization-related representations. By reusing the backbone trained from the source dataset and mainly optimizing the extractor for new datasets, the proposed method can effectively transfer localization knowledge across different dataset environments.

Experimental results demonstrated that the shared-backbone strategy achieves faster convergence and lower testing RMSE in most adaptive learning scenarios compared with training the entire model from scratch. The results also showed that the reused backbone provides beneficial prior localization knowledge, enabling the model to adapt more efficiently to new physical configurations. Furthermore, the training time comparison verified that the proposed framework reduces the computational cost of adaptation while maintaining effective localization performance. Overall, the proposed method provides an efficient and scalable solution for RSS-based indoor localization in dynamic and heterogeneous environments.

In practical deployment scenarios, the proposed lightweight adaptation strategy is particularly beneficial for environments requiring frequent recalibration or rapid deployment under limited computational resources. Although the absolute reduction in training time may appear moderate for a single adaptation process, the cumulative computational savings become increasingly important when the model is repeatedly updated across multiple environments or edge devices with constrained hardware capability.

Nevertheless, several limitations still exist in the current work. The proposed framework is mainly evaluated on relatively structured RSS localization datasets, and the fully connected architecture may have limited representation capability in more complicated environments involving large-scale buildings, severe environmental dynamics, or dense obstacles. In addition, although the shared-backbone strategy improves adaptation efficiency, the scalability of the proposed method under significantly larger domain discrepancies still requires further investigation.

Future work will focus on extending the proposed framework toward more complicated localization scenarios, including dynamic environments and large-scale multi-floor deployments. More advanced architectures, such as attention-based or hybrid lightweight models, as well as data augmentation and self-supervised adaptation strategies, will also be investigated to further improve generalization capability and robustness.

\bibliographystyle{IEEEtran}
\bibliography{ref}

@article{lin2026variational,
  title={Variational Bayesian Multi-Source Localization in Complex Multipath Environments},
  author={Lin, Zhipeng and Cai, Xuezhao and He, Yunhong and Guo, Lantu and Xu, Yongjie and Wei, Ni and Zhu, Qiuming and Wu, Qihui},
  journal={IEEE Transactions on Communications},
  volume={74},
  pages={7090--7105},
  year={2026},
  publisher={IEEE}
}

@ARTICLE{7366598,
  author={Liu, Chen and Fang, Dingyi and Yang, Zhe and Jiang, Hongbo and Chen, Xiaojiang and Wang, Wei and Xing, Tianzhang and Cai, Lin},
  journal={IEEE Transactions on Wireless Communications}, 
  title={RSS Distribution-Based Passive Localization and Its Application in Sensor Networks}, 
  year={2016},
  volume={15},
  number={4},
  pages={2883-2895},
  doi={10.1109/TWC.2015.2512861}}

@ARTICLE{11362365,
  author={Yuli Martin Adiyatma, Farid and Cherntanomwong, Panarat and Joko Suroso, Dwi},
  journal={IEEE Open Journal of the Communications Society}, 
  title={Dynamic Ensemble Learning With Received Signal Strength Transformation for Robust Multi-Floor Wi-Fi Indoor Localization}, 
  year={2026},
  volume={7},
  number={},
  pages={978-996},
  doi={10.1109/OJCOMS.2026.3657332}}

@ARTICLE{8030076,
  author={Liu, Kai and Zhang, Hao and Ng, Joseph Kee-Yin and Xia, Yusheng and Feng, Liang and Lee, Victor C. S. and Son, Sang H.},
  journal={IEEE Transactions on Industrial Informatics}, 
  title={Toward Low-Overhead Fingerprint-Based Indoor Localization via Transfer Learning: Design, Implementation, and Evaluation}, 
  year={2018},
  volume={14},
  number={3},
  pages={898-908},
  doi={10.1109/TII.2017.2750240}}

@article{park2022multidirectional,
  title={Multidirectional differential RSS technique for indoor vehicle navigation},
  author={Park, Pangun and Di Marco, Piergiuseppe and Jung, Mingyu and Santucci, Fortunato and Sung, Tae Kyung},
  journal={IEEE Internet of Things Journal},
  volume={10},
  number={1},
  pages={241--253},
  year={2022},
  publisher={IEEE}
}

@article{yin2017distributed,
  title={Distributed recursive Gaussian processes for RSS map applied to target tracking},
  author={Yin, Feng and Gunnarsson, Fredrik},
  journal={IEEE Journal of Selected Topics in Signal Processing},
  volume={11},
  number={3},
  pages={492--503},
  year={2017},
  publisher={IEEE}
}

@ARTICLE{9542942,
  author={He, Jiajun and Chun, Young Jin and So, Hing Cheung},
  journal={IEEE Internet of Things Journal}, 
  title={A Unified Analytical Framework for RSS-Based Localization Systems}, 
  year={2022},
  volume={9},
  number={9},
  pages={6506-6519},
  doi={10.1109/JIOT.2021.3114232}}

@ARTICLE{10476652,
  author={Wang, Chao and He, Jiajun and Xiong, Wenxin and So, Hing Cheung and Chun, Young Jin},
  journal={IEEE Transactions on Aerospace and Electronic Systems}, 
  title={Received Signal Strength Characterization in Mixed LOS and NLOS Environments}, 
  year={2024},
  volume={60},
  number={4},
  pages={5613-5620},
  doi={10.1109/TAES.2024.3379483}}

@ARTICLE{11165772,
  author={He, Jiajun and Quoc Ngo, Hien and Wang, Chao and Yin, Feng and Cheung So, Hing and Shin, Hyundong and Matthaiou, Michail},
  journal={IEEE Transactions on Wireless Communications}, 
  title={RSS Localization in Cell-Free Massive MIMO: Algorithms, Analysis, and Implementation}, 
  year={2026},
  volume={25},
  number={},
  pages={4021-4037},
  doi={10.1109/TWC.2025.3607721}}

@ARTICLE{11458632,
  author={Zhang, Jiaming and He, Jiajun and Lu, Tianyu and Zhang, Jie and Yurduseven, Okan},
  journal={IEEE Antennas and Wireless Propagation Letters}, 
  title={Adversarial Learning-Based Radio Map Reconstruction for Fingerprinting Localization}, 
  year={2026},
  volume={},
  number={},
  pages={1-5},
  doi={10.1109/LAWP.2026.3679273}}

@inproceedings{dichasus2021,
	author    = {Florian Euchner and Marc Gauger and Sebastian D\"orner and Stephan ten Brink},
	title     = {{A Distributed Massive MIMO Channel Sounder for "Big CSI Data"-driven Machine Learning}},
	booktitle = {WSA 2021; 25th International ITG Workshop on Smart Antennas},
	year      = {2021}
}

@inproceedings{feng2022analysis, 
  title={An analysis of the properties and the performance of wifi rtt for indoor positioning in non-line-of-sight environments},
  author={Feng, Xu and others},
  booktitle={\textit{Int. Conf. Location Based Services (LBS), 17th}},
  year={2022}
}

@ARTICLE{11016042,
  author={Xiang, Tiange and others},
  journal={\textit{IEEE Internet Things J.}}, 
  title={MoDe{FA}: Multi-Observer and Denoising-Enhanced Fingerprint Augmentation for Semi-Supervised {W}i-{F}i {RSS}-Based Indoor Positioning}, 
  year={2025},
  volume={},
  number={},
  pages={1-1}}

@article{yuen2022wi,
  title={Wi-{F}i and Bluetooth contact tracing without user intervention},
  author={Yuen, Brosnan and others},
  journal={\textit{IEEE Access}},
  volume={10},
  pages={91027--91044},
  year={2022},
  publisher={IEEE}
}

@ARTICLE{11397142,
  author={Cheng, Xin and He, Yu and Li, Menglu and Li, Ruoguang and Shu, Feng and Han, Guangjie},
  journal={IEEE Transactions on Wireless Communications}, 
  title={A Fingerprint Database Generation Method for RIS-Assisted Indoor Positioning}, 
  year={2026},
  volume={25},
  number={},
  pages={11932-11948},
  doi={10.1109/TWC.2026.3662428}}

@article{so2011linear,
  title={Linear least squares approach for accurate received signal strength based source localization},
  author={So, Hing Cheung and Lin, Lanxin},
  journal={IEEE Transactions on signal processing},
  volume={59},
  number={8},
  pages={4035--4040},
  year={2011},
  publisher={IEEE}
}

@article{he2026rss,
  title={RSS-based Localization with a Single Receiver: Method and Stochastic Analysis},
  author={He, Jiajun and Ho, KC and Ngo, Hien Quoc and Wang, Chao and Yu, Han and So, Hing Cheung and Shin, Hyundong and Matthaiou, Michail},
  journal={IEEE Transactions on Wireless Communications},
  year={2026},
  publisher={IEEE}
}

@ARTICLE{10274764,
  author={Gao, Jun and Wu, Dongze and Yin, Feng and Kong, Qinglei and Xu, Lexi and Cui, Shuguang},
  journal={IEEE Journal on Selected Areas in Communications}, 
  title={MetaLoc: Learning to Learn Wireless Localization}, 
  year={2023},
  volume={41},
  number={12},
  pages={3831-3847},
  doi={10.1109/JSAC.2023.3322766}}

@ARTICLE{9122501,
  author={He, Jiajun and So, Hing Cheung},
  journal={IEEE Sensors Journal}, 
  title={A Hybrid TDOA-Fingerprinting-Based Localization System for LTE Network}, 
  year={2020},
  volume={20},
  number={22},
  pages={13653-13665},
  doi={10.1109/JSEN.2020.3004179}}

@article{zhang2025fingan,
  title={FinGAN: An Interpretable RSS Generation Network for Scalable Fingerprint Localization},
  author={Zhang, Jiaming and He, Jiajun and Zhang, Jie and Yurduseven, Okan},
  journal={arXiv preprint arXiv:2509.26286},
  year={2025}
}

@ARTICLE{10077563,
  author={Zhang, Bangjie and Xu, Gang and Yu, Hanwen and Wang, Hui and Pei, Hao and Hong, Wei},
  journal={IEEE Transactions on Geoscience and Remote Sensing}, 
  title={Array 3-D SAR Tomography Using Robust Gridless Compressed Sensing}, 
  year={2023},
  volume={61},
  number={},
  pages={1-13},
  doi={10.1109/TGRS.2023.3259980}}

@ARTICLE{11026791,
  author={Foliadis, Anastasios and Castañeda Garcia, Mario H. and Stirling-Gallacher, Richard A. and Thomä, Reiner S.},
  journal={IEEE Transactions on Wireless Communications}, 
  title={Transfer Learning for CSI-Based Positioning With Multi-Environment Meta-Learning}, 
  year={2025},
  volume={24},
  number={11},
  pages={9735-9748},
  doi={10.1109/TWC.2025.3575001}}

\end{document}